\documentclass[11pt]{article}

\usepackage{fix-cm}
\usepackage{graphicx}
\usepackage[hyphens]{url}
\usepackage{xcolor}
\definecolor{darkgreen}{rgb}{0.0, 0.5, 0.0}

\usepackage{natbib}
\setcitestyle{authoryear}

\usepackage{tabularx}
\usepackage{multirow}
\usepackage{lscape}
\usepackage{longtable}
\usepackage{xltabular}
\usepackage{float}
\usepackage{booktabs}

\usepackage{nameref}

\title{The Role of Team Diversity in AI Systems Development}

\author{
Ronnie de Souza Santos\thanks{University of Calgary, Calgary, Canada. Email: ronnie.desouzasantos@ucalgary.ca}
\and
Maria Teresa Baldassarre\thanks{Università di Bari, Bari, Italy. Email: mariateresa.baldassarre@uniba.it}
\and
Cleyton Magalhaes\thanks{Universidade Federal Rural de Pernambuco, Recife, Brazil. Email: cleyton.vanut@ufrpe.br}
}

\date{}

\begin{document}

\maketitle

\maketitle

\begin{abstract}
The widespread integration of AI technologies has intensified concerns about fairness and bias, as these systems often perpetuate societal inequalities through flawed data and design choices. While software engineering research has largely concentrated on technical solutions, such as improving datasets and models, the social dynamics that shape AI outcomes remain underexplored. This study investigates the role of team diversity in the development of AI systems. Drawing from the experience of four AI-focused teams working in a large software company operating in Brazil and Portugal, and collaborating with global clients, the study explores how diverse teams influence the development of AI systems. Using Grounded Theory, we conducted 25 interviews with software professionals involved in projects spanning domains such as education, energy, accessibility, and facial recognition. Although our study is conducted in an organizational setting, the variety of projects, from regional to multinational, ensures exposure to global development practices and diverse team dynamics, bringing a variety of perspectives into our findings. Our analysis revealed six key roles that team diversity played in AI development: diversifying perspectives for bias identification, bringing empathy to AI development, addressing systemic discrimination, supporting inclusive and participatory decision-making, using diversity as a safeguard against bias, and fostering broadened thinking in problem-solving. These findings highlight the importance of incorporating diverse perspectives in AI projects and offer practical recommendations for integrating fairness considerations into software development practices.

\end{abstract}

\section{Introduction}\label{sec:introduction}

Software systems permeate nearly every aspect of modern society, from education and healthcare to daily services and leisure~\citep{albusays2021diversity}. As machine learning and artificial intelligence are increasingly adopted, discussions on fairness and bias in these applications have gained significant attention. While these systems have the potential to transform industries, they also risk embedding systemic problems that can exacerbate discrimination and reinforce existing social inequalities ~\citep{brun2018software, zhang2021ignorance, galhotra2017fairness}. Recent reports in various domains have shown how the integration of AI into critical decision-making processes has increased the visibility of biases, drawing attention to how these systems may unintentionally perpetuate inequalities~\citep{fountain2022moon, owens2020those, lee2018detecting, ryan2021digital, wu2020gender, rodriguez2023lgbtq}.

Bias in AI systems often stem from systematic errors in decision-making in the software development processes, usually due to training data reflecting historical inequities. Consequently, AI models replicate and reinforce harmful patterns, leading to discriminatory outcomes~\citep{paez2021negligent, kelly2023algorithmic}. For example, black people have faced higher insurance rates, lower credit scores, and harsher outcomes in criminal justice due to biased algorithms~\citep{fountain2022moon, owens2020those, garcia2024algorithmic}. Gender discrimination in hiring processes has been exacerbated by algorithms that favor male candidates in technical roles, while LGBTQIA+ communities often encounter biased content moderation algorithms that suppress their content~\citep{wu2020gender, rodriguez2023lgbtq}. Immigrants and minorities experience bias in facial recognition systems, which frequently misidentify individuals with darker skin tones, resulting in increased surveillance and wrongful accusations~\citep{limante2024bias, laupman2022biased}.

In this context, the concept of software fairness debt~\citep{desoftware} has been introduced to highlight various sources of bias (e.g., cognitive, design, historical, societal, and technical) which, if left unaddressed, undermine fairness and lead to harmful societal impacts. Following this, while traditional software engineering research has focused primarily on technical solutions such as datasets, models, and algorithms~\citep{serna2022sensitive, nguyen2024literature}, recent discussions emphasize the importance of integrating social factors across the AI development lifecycle. These factors that arise from developers’ cognitive processes, design decisions, and societal influences interact with technical elements and shape both the outcomes of AI systems and their broader implications.

Considering this scenario, this study investigates the role of team diversity in AI development, with a focus on how it contributes to identifying and mitigating biases. By researching real-world environments, we analyzed how team diversity influences the development of fairer and unbiased AI systems and supports the prevention of harms that may arise when biases remain unaddressed. The central research question driving this study is: \textbf{\textit{What is the role of team diversity in the identification and mitigation of biases in AI system development?}} By answering this question, this paper presents four key contributions addressing significant gaps in what is known about this topic:

\begin{enumerate}
    \item \textbf{Exploring the influence of team diversity in AI development:} Our study identifies six roles of team diversity in AI development, showing how team composition can support the recognition and mitigation of biases. \\
    
    \item \textbf{Enhancing understanding of social implications in AI development:} Our findings highlight how diverse teams are better positioned to consider the broader effects of AI systems, helping to reduce risks of unfair practices or outcomes. \\
    
    \item \textbf{Safeguarding against algorithmic biases:} Our collected evidence indicates that team diversity might help reduce algorithmic biases by introducing perspectives that question assumptions and reveal hidden issues in datasets and models. \\
    
    \item \textbf{Bridging technical and societal goals in AI development:} Our study illustrates how diverse teams contribute to aligning AI development with societal objectives such as fairness, equity, and inclusivity, thereby supporting the creation of systems that serve a wider range of users. \\
\end{enumerate}

The structure of this paper is as follows: Section~2 provides background on algorithmic bias and software team diversity. Section~3 details the research methodology. Section~4 presents the thematic analysis results. Section~5 discusses the findings and their implications, and Section~6 concludes with insights and future research directions.

\section{Background} \label{back}

The two central topics of this research (bias and discrimination in AI systems and software team dynamics) are not typically part of the same discussions within software engineering. Therefore, this section introduces key concepts related to fairness in software systems, algorithmic bias, societal impacts, and team diversity to support a better understanding of the research problem and enable comparisons with our findings.

\subsection{Bias and Discrimination in AI}

Fueled by vast amounts of data generated through everyday online activities, AI systems have been promoted for their potential to enhance efficiency and objectivity~\citep{ferrara2023fairness, lee2020artificial}. However, biases in the data used to train these systems can lead to discriminatory outcomes, as AI tends to replicate societal inequalities reflected in that data~\citep{allen2020artificial, adams2020diversity}. These biases can disproportionately affect marginalized groups, causing harm and reinforcing pre-existing social inequalities. This phenomenon is defined as \textit{algorithmic discrimination}~\citep{bigman2023algorithmic, dehal2024exposing, santos2023perspective, schwarting2022organization} and refers to issues arising from biased software systems that disproportionately affect marginalized groups based on attributes such as ethnicity, race, gender, sexual orientation, disability, and socioeconomic status.

To address these challenges, the concept of \textit{software fairness} has emerged, emphasizing the mitigation of biases to promote equitable outcomes. Software fairness is the practice of designing and deploying computational systems in ways that remain inclusive across demographic groups, regardless of attributes such as race, gender, ethnicity, or socioeconomic status~\citep{galhotra2017fairness, verma2018fairness, starke2022fairness, chen2024fairness, ferrara2023fairness}. Within software engineering, fairness is associated with preventing discrimination, fostering inclusivity, and reducing biases embedded in datasets and development processes~\citep{brun2018software, zhang2021ignorance}. If biases are left unaddressed, they can result in significant risks and potential harm to individuals and communities~\citep{hort2024bias}. In the context of AI, such biases may arise from multiple sources~\citep{desoftware}:

\begin{itemize}
    \item \textit{Cognitive bias}: Developers’ individual judgments, assumptions, or attitudes may influence system behavior.  
    \item \textit{Design bias}: Decisions in interface or feature design may unintentionally privilege certain demographic groups.  
    \item \textit{Historical bias}: Reliance on legacy data can reproduce patterns of past discrimination.  
    \item \textit{Model bias}: Limitations or errors in model architecture and training processes can distort outcomes.  
    \item \textit{Requirements bias}: Incomplete or biased requirements may constrain inclusivity in system functionality.  
    \item \textit{Societal bias}: Unfair social structures and norms can become embedded in technological systems.  
    \item \textit{Testing bias}: Insufficiently representative testing practices may overlook the needs of some user groups.  
    \item \textit{Training bias}: Training datasets with unbalanced or skewed representations can lead to unfair or biased performance across groups.  
\end{itemize}

In practice, implementing fairness-oriented strategies in software development is essential for mitigating risks, promoting equitable outcomes, and preventing the accumulation of \textit{software fairness deb}~\citep{desoftware}. As bias can arise from multiple domains, including human judgment, data, design choices, and societal structures, fairness must be addressed comprehensively across the entire development process.

\subsection{Software Fairness}

In artificial intelligence, fairness refers to the extent to which systems treat users equitably, without discrimination based on protected attributes such as race, gender, or age~\citep{brun2018software, chen2024fairness}. Currently, fairness is recognized as a non-functional requirement in software, alongside reliability, performance, and security~\citep{brun2018software}. The literature often distinguishes between two categories: \textit{individual fairness}, which requires similar treatment for similar individuals, and \textit{group fairness}, which aims to achieve equitable outcomes across demographic groups~\citep{chen2024fairness, soremekun2022software}. These categories frame much of the technical work on fairness in software systems.

Most technical approaches address fairness by focusing on models or datasets, frequently at the individual or group levels. At the individual level, strategies include fairness through unawareness, which removes protected attributes from the data or model; fairness through awareness, which incorporates these attributes to ensure comparable treatment; counterfactual fairness, which evaluates whether outcomes remain consistent if protected attributes were hypothetically altered; and causal fairness, which analyzes relationships among variables to determine whether sensitive attributes influence results~\citep{zhang2021ignorance, chen2024fairness, soremekun2022software}. At the group level, approaches typically rely on statistical criteria such as statistical parity, which requires equal distribution of positive outcomes across groups; equalized odds, which ensures comparable error rates; and equal opportunity, which balances true positive rates across groups~\citep{chen2024fairness, soremekun2022software}. Each of these techniques involves trade-offs, since achieving fairness under one definition may reduce it under another and can also affect model accuracy, making the choice of method highly dependent on the application domain, project priorities, and system constraints.

Beyond models and datasets, researchers in software engineering have argued for embedding fairness practices throughout the development lifecycle~\citep{brun2018software, soremekun2022software}. This includes fairness-aware requirements engineering~\citep{zhang2021ignorance}, which identifies stakeholder values and potential sources of bias early, as well as fairness testing~\citep{galhotra2017fairness, chen2024fairness}, which evaluates system outputs for discriminatory behavior. Recent studies also highlight the importance of preserving fairness during system evolution through regression testing and continuous monitoring, ensuring fairness properties are maintained across software updates~\citep{soremekun2022software}.

\subsection{Fairness and the Effects of Bias in AI Systems in Society}
Fairness in AI systems is ultimately determined by how biases are identified and addressed before deployment. When such biases remain unmitigated, fairness is compromised, leading to social consequences such as discrimination, inequality, and diminished trust in technology. Prior work has mapped different forms of bias in AI and discussed several harmful effects that arise when fairness is neglected in software development~\citep{desoftware}, as summarized below:

\begin{itemize}
\item \textit{Exacerbation of social inequality}: Data and algorithms risk reproducing biases against historically disadvantaged populations~\citep{hoffmann2019fairness}. The negligence of software fairness amplifies disproportions by benefiting specific groups or reinforcing historical biases, thereby widening social divides and deepening inequalities~\citep{desoftware}. \\

\item \textit{Legal concerns}: Disregarding fairness raises legal issues related to discrimination and privacy, as well as ongoing debates over regulatory standards and accountability. For instance, research on fairness in machine learning algorithms has been shown to be strongly influenced by U.S. legal frameworks~\citep{kirat2023fairness}. \\  

\item \textit{Limited algorithmic reliability}: Ignoring fairness reduces the reliability of algorithms, resulting in inconsistencies and errors that diminish user confidence in automated decision-making. For example, a sentiment analyzer has been shown to link negative sentiment to “man” and positive sentiment to “woman”~\citep{soremekun2022software}. \\  

\item \textit{Proliferation of discrimination}: The absence of fairness sustains discriminatory practices, producing unfair treatment based on factors such as race or gender. Documented cases include Black individuals receiving lower credit scores or higher delivery fees due to algorithmic discrimination~\citep{fountain2022moon, brun2018software}. \\  

\item \textit{Psychological harms}: Insufficient fairness exposes users to bias and unfair treatment, leading to emotional distress, anxiety, and other negative impacts on mental health and quality of life~\citep{bigman2023algorithmic}. \\  

\item \textit{Reduced algorithmic accuracy}: A lack of fairness compromises accuracy, allowing biases to distort decisions and generate outcomes that deviate from intended goals. For instance, translations from gender-neutral languages may reproduce societal biases~\citep{brun2018software}. \\  

\item \textit{Reinforcement of stereotypes}: Failure to address fairness perpetuates harmful stereotypes, obstructs diversity and inclusion, and contributes to systemic discrimination. A well-known example is facial recognition systems performing poorly on African-American faces~\citep{fountain2022moon, brun2018software}. \\  

\item \textit{Weakening of justice}: Neglecting fairness undermines justice by embedding biases into decision-making processes, infringing on fundamental rights, and reducing trust in institutions that rely on AI ~\citep{pfeiffer2023algorithmic, giannopoulos2024fairness}.
\end{itemize}

Given the scale and impact of these consequences, fairness in AI software cannot be ensured through technical interventions alone. Social and organizational factors, particularly the individuals involved in the development process, play a critical role in shaping outcomes. Prior research in software engineering has shown that the composition of development teams influences how software solutions are conceived and implemented~\citep{rodriguez2021perceived}, yet it has not explored in depth how fairness considerations are incorporated into these systems through diverse perspectives.

\subsection{Software Team Diversity}
Team diversity encompasses the range of individual differences within groups and is typically categorized into three forms: value diversity, which refers to differences in beliefs, goals, and guiding principles among team members; informational diversity, which captures variation in professional experience, technical knowledge, and educational background; and social diversity, which reflects attributes such as gender, age, race, and ethnicity~\citep{rodriguez2021perceived}. Beyond these demographic and experiential factors, diversity in software engineering also extends to cognitive diversity, which reflects how individuals process information, apply reasoning strategies, and approach problem-solving tasks~\citep{menezes2018diversity}. Collectively, these dimensions enable software teams to address complex challenges from multiple angles, often resulting in more innovative, adaptive, and resilient solutions.

Empirical research shows that heterogeneous teams often outperform homogeneous ones in productivity, creativity, and problem-solving~\citep{pieterse2006software, gila2014impact}. For example, groups with varied skills and experiences are better positioned to identify blind spots, consider alternative perspectives, and adapt to evolving project demands. At the same time, diversity can introduce challenges such as communication barriers, cultural misunderstandings, and interpersonal conflict, which may slow decision-making and weaken cohesion~\citep{wickramasinghe2015diversity, verwijs2023double}. These risks emphasize the importance of effective leadership and inclusive practices to harness the benefits of diversity while mitigating potential drawbacks~\citep{verwijs2023double}.

In the context of software engineering for AI, the role of team diversity remains underexplored. Evidence from the domains of organizational behavior and ethics suggests that diverse teams can improve transparency, facilitate the identification of biases, and enhance the trustworthiness of AI systems by incorporating varied perspectives~\citep{boch2022ethical}. However, despite these indications, little empirical work has examined how team diversity specifically shapes fairness considerations in AI software development. Addressing this gap is key for understanding how team diversity can be leveraged to promote fairness and inclusivity in technological outcomes~\citep{desoftware, nguyen2024literature}.

\section{Method} \label{method}
We conducted this study using Grounded Theory (GT), an inductive research methodology widely adopted for exploring complex social and technical phenomena \citep{charmaz2014constructing, adolph2011using, stol2016grounded}. This approach enabled us to analyze empirical data collected from software professionals involved in AI system development. Below, we provide a comprehensive overview of the study's context, data collection, data analysis, auditing, and ethical considerations. These sections summarize our research methodology, which adheres to established guidelines for grounded theory and its application in software engineering \citep{charmaz2014constructing, adolph2011using, stol2016grounded, ralph2020empirical}.

While we acknowledge that alternative approaches such as Social-Technical Grounded Theory are well-suited to studies that explicitly aim to model the interplay between social and technical systems, we chose to follow the classical Grounded Theory approach as described by Charmaz \citep{charmaz2014constructing} and adapted for software engineering \citep{adolph2011using, stol2016grounded, ralph2020empirical} because our objective is not to construct a formal theory, but to explore how team diversity influences AI development through a flexible, inductive process. This approach allowed key concepts to emerge organically from participants’ experiences, without relying on a predefined socio-technical framework.

\begin{figure}[H]
\centering
\includegraphics[width=0.7\linewidth]{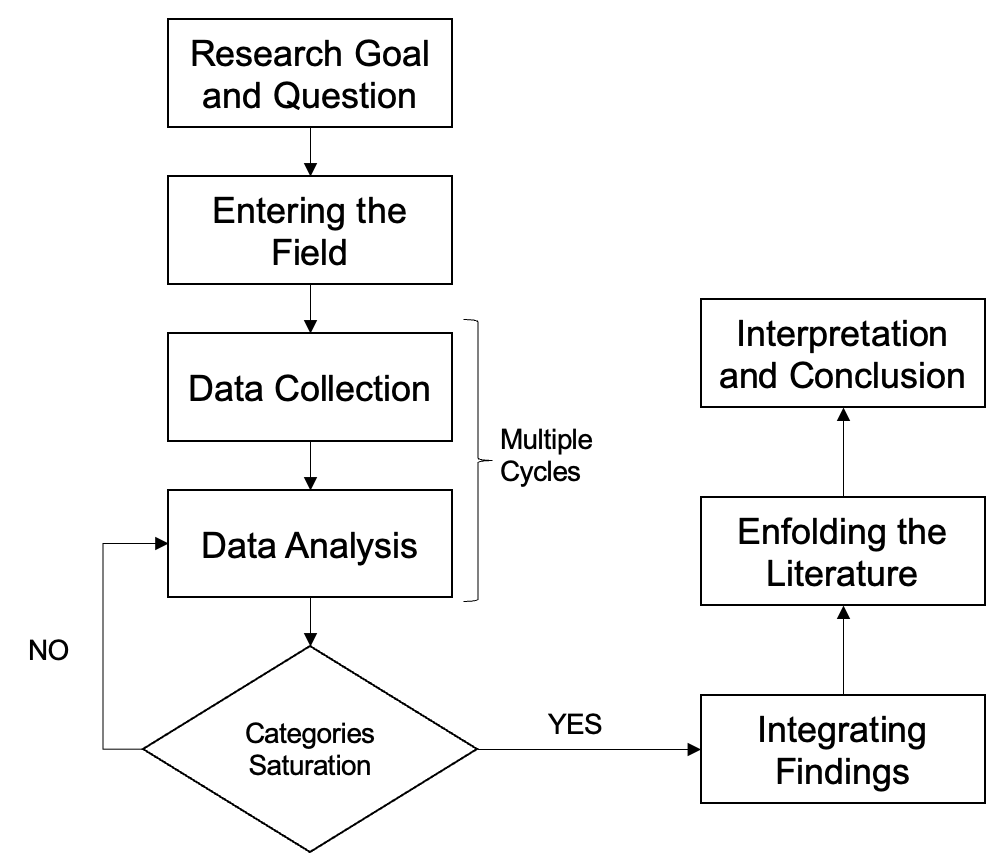}
\vspace{-0.1in}
\caption{Grounded Theory Process Used in This Study.}
\label{fig:gt}
\end{figure}

Hence, in line with established guidelines \citep{charmaz2014constructing, adolph2011using, stol2016grounded}, our research process involved the following steps (illustrated in Figure \ref{fig:gt}):

\begin{enumerate}
    \item \textbf{Research goal and question:} We defined our central research question focusing on how team diversity influences the development of AI systems.
    
    \item \textbf{Entering the field:} We selected four AI projects within a large software company and identified initial participants using convenience and snowball sampling.
    
    \item \textbf{Data collection:} We conducted cycles of semi-structured interviews, having 25 participants distributed across three iterative rounds, and supplemented these with three open-ended questionnaires to capture additional perspectives.
    
    \item \textbf{Data analysis:} Using line-by-line, focused, and theoretical coding techniques, we analyzed transcripts after each round to identify emerging concepts, refine categories, and guide theoretical sampling.
    
    \item \textbf{Category saturation check:} We assessed saturation throughout the process and concluded data collection once additional interviews no longer yielded new concepts.
    
    \item \textbf{Integrating findings:} We grouped the resulting categories into higher-level themes that explain how diversity contributed to AI development.
    
    \item \textbf{Enfolding the literature:} After category integration, we revisited the literature to compare and contrast our findings with existing work on software fairness and team diversity.
    
    \item \textbf{Interpretation and conclusion:} We synthesized the findings into a cohesive interpretation, identifying practical implications for team diversity in AI development.
\end{enumerate}

In summary, our process unfolded across three main phases, preparation, iterative data collection, and analysis and interpretation, with each phase contributing to a deeper understanding of how team diversity influences AI development. Below, we provide details on each step.

\subsection{Software Development Context}
This study was conducted in the context of four AI-focused projects under development within a large software company founded in 1996, with an operational presence in Brazil and Portugal. The company provides solutions across sectors such as finance, telecommunications, government, manufacturing, services, and utilities. Of the company’s 1,200 employees, over 70\% are directly engaged in software development, organized into more than 70 teams. These teams comprise professionals with diverse technical expertise (e.g., programmers, quality assurance engineers, designers) and personal backgrounds (e.g., gender and ethnic diversity). They are experienced in various development methodologies, such as Scrum, Kanban, and Waterfall, and deliver solutions to clients across North America, Latin America, Europe, and Asia.

From this context, we identified and selected four AI-focused projects that could provide the data necessary for our study. This selection followed a convenience sampling approach, as we selected projects based on the availability of participants to be involved in the study and the company’s willingness to allow professionals to discuss the projects. Although all participants are affiliated with the same company, the unit of analysis in this study is the individual software project under development—not the organization as a whole. Each project was carried out for a different client and involved distinct goals, requirements, data contexts, and collaboration dynamics. Importantly, these projects were co-developed in close partnership with professionals from the client organizations, ensuring that development practices reflected global standards and incorporated cross-organizational perspectives. The projects under investigation were:

\begin{itemize}
    \item \textbf{Project A} was developed in collaboration with a multinational technology company headquartered in China, with operations in over 180 countries. The project focused on real-time sign language translation using deep learning (this project is reported in \citep{mari2020libras}).

    \item \textbf{Project B} was linked to one of the largest state-controlled oil and gas companies in the world, which holds a significant presence in Latin America’s energy sector. The project centered on the development of digital twins and predictive models.

    \item \textbf{Project C} involved a regional educational initiative, exploring the application of large language models (LLMs) in academic settings.

    \item \textbf{Project D} was conducted in partnership with one of the largest beauty and cosmetics companies in Brazil, a market leader in Latin America, and focused on computer vision-based facial recognition systems (this project is reported in \citep{botocario}).
\end{itemize}

Due to contractual restrictions, more detailed information about Projects B and C is not available, different from Project A and D, which were released, and a report is available online \citep {mari2020libras, botocario}. Despite being situated within a single organizational context, the combination of multinational and regional clients, global collaboration, and domain-specific diversity across these four projects contributes to a rich, transferable understanding of how team diversity influences AI development.

\subsection{Data Collection} \label{collection}
Participants in this study were software professionals actively engaged in the selected projects, spanning roles such as designers, software engineers, data scientists, and testers. To initiate participant recruitment, we employed convenience sampling \citep{baltes2022sampling}, beginning with professionals from Project A and subsequently expanding to Projects B, C, and D. An open invitation was sent to these professionals, inviting them to participate in the study and asking them to provide their preferred dates and times for interviews. Therefore, consistent with the convenience sampling strategy, participants were selected based on their availability and willingness to contribute to the research, which was particularly important given the varying schedules and responsibilities within the software teams. 

Following this initial recruitment, we applied snowball sampling, wherein participants referred colleagues they believed had relevant expertise or unique perspectives on the projects to contribute to the research. This strategy allowed us to extend our reach to professionals who might not have responded to the initial invitation but were well-positioned to contribute with valuable experiences. For instance, participants from early interviews suggested involving team members in roles such as quality assurance or user experience design, who provided complementary perspectives to those working on data science and coding activities.

Additionally, to further refine our sample, as required by GT, we incorporated theoretical sampling \citep{charmaz2014constructing} by sending direct invitations to individuals who could provide valuable insights into phenomena identified but not yet fully understood. This involved targeting participants with specific levels of experience, such as senior developers or testers, and those from diverse demographic backgrounds, including underrepresented groups in software engineering. For instance, as data began to emerge, we identified the need for insights from professionals working in leadership roles to capture perspectives on less technical themes, such as team dynamics and soft skills, which had surfaced during earlier interviews. We also sought participants from projects involving varying levels of sensitivity in user data, such as those working on applications directly impacting individuals versus industrial systems. This strategy ensured that our sample reflected a broad spectrum of expertise and project contexts, enabling a deeper exploration of team diversity's role in bias identification and mitigation.

In general, our data collection strategy relied on semi-structured interviews conducted in three iterative rounds between June 1 and July 5, 2024. The main interview questions are presented in Table \ref{tab:InterviewScript}. Twenty-five professionals contributed, and the duration of the interviews ranged from 23 to 42 minutes. The interviews generated 4 hours and 49 minutes of audio recordings and 504 pages of transcribed data. Additionally, memos were written during and after interviews, capturing contextual details and initial impressions that supported the iterative refinement of data collection and analysis. Early rounds of interviews focused on gathering general perspectives on bias in AI development. As the study progressed, subsequent rounds focused on more specific topics, including team diversity and the concept of fairness debt. To accommodate three key participants who were unable to attend live interviews, we provided open-ended questionnaires aligned with the interview script. These questionnaires ensured that their insights were captured consistently with the broader data collection process, maintaining the depth and relevance of their contributions. 

The iterative nature of GT allowed the interview guide to evolve based on emerging insights, such as shifting focus from highly technical aspects to managerial characteristics of the problem as they became prominent. Following the concept of theoretical saturation \citep{glaser1978theoretical}, data collection concluded when no new insights, codes, or themes emerged from the interviews. This saturation point is critical in Grounded Theory, as it signals that the data sufficiently supports the development of robust theoretical categories. In our study, theoretical saturation was reached by the third round of interviews, during which recurring themes, such as empathy and the value of diverse perspectives, were the only ones to continue to appear in participants’ narratives. 

For instance, as later interviews reiterated the importance of interpersonal dynamics and team composition in identifying and mitigating biases, no additional dimensions or subcategories of these themes were identified. This indicated that the data collection captured the range of experiences relevant to the research focus. To ensure rigor, we monitored saturation closely using memos to track the emergence and repetition of themes. Additionally, the decision to end data collection was collectively reviewed and agreed upon by the research team, ensuring that saturation was not prematurely declared but grounded in a systematic and iterative engagement with the data.

\subsection{Data Analysis} \label{analysis}
Data analysis adhered to the core principles of grounded theory \citep{charmaz2014constructing}, beginning with an in-depth review of the interview transcripts. These transcripts, derived from participant interviews, formed the foundation for the analysis, capturing rich narratives on team diversity and bias in AI development. Analysis commenced with line-by-line coding, based on the examination of the transcripts to identify initial concepts embedded in participants’ narratives. This process generated a broad set of codes that directly reflected participants' lived experiences. 

For instance, codes such as \textit{empathy as a soft skill} and \textit{context sensitivity} were derived from specific participant statements and later evolved into the broader category. To ensure rigor, we employed constant comparison \citep{glaser1978theoretical}, a systematic approach where each new segment of data was continuously compared to existing codes. This method enabled us to refine categories and validate emerging patterns, ensuring that our analysis remained grounded in the data. As recurring codes were identified, relationships between them began to surface, guiding the transition from line-by-line coding to focused coding. 

\begin{figure}[H]
\centering
\includegraphics[width=0.95\linewidth]{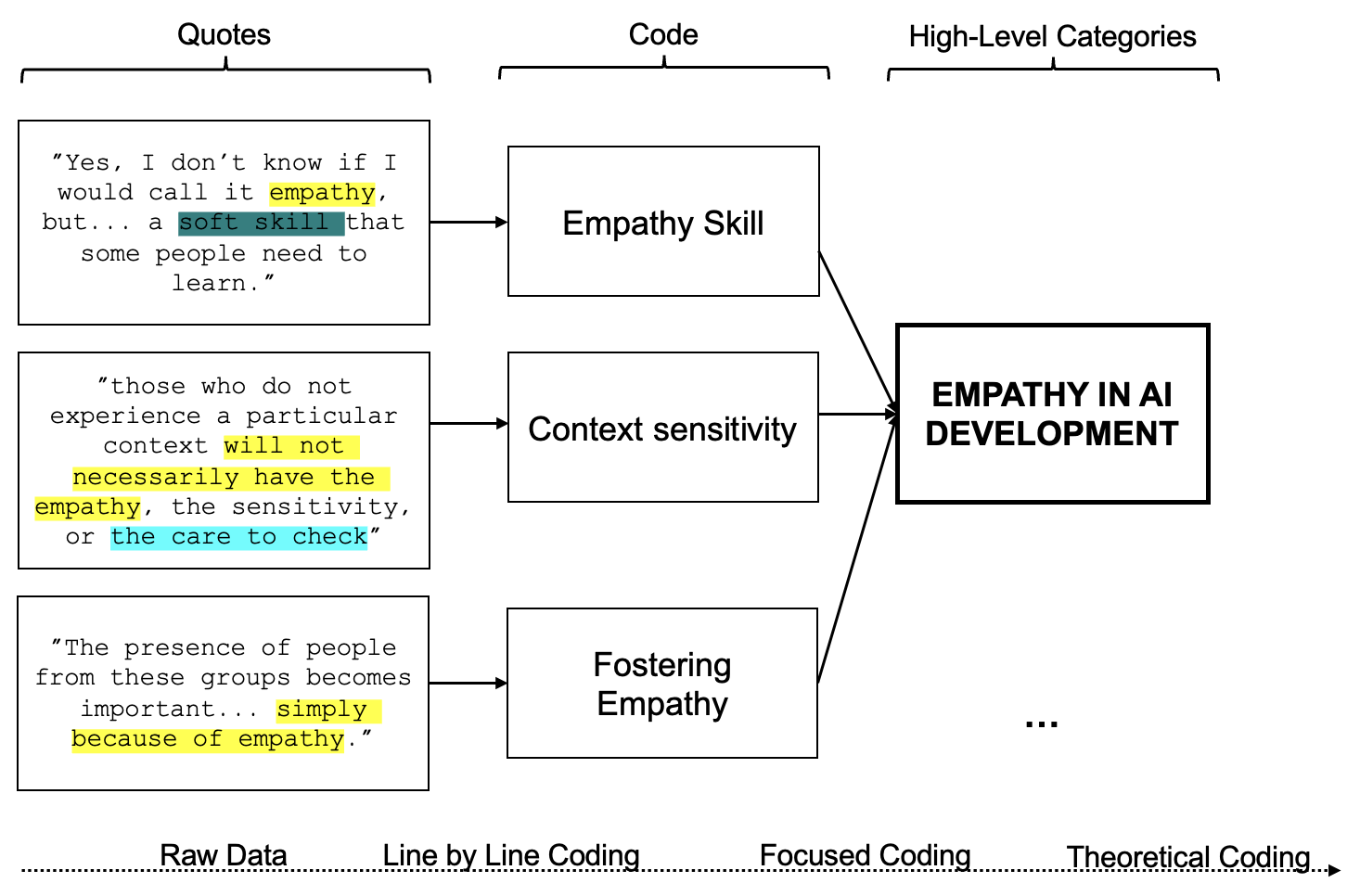}
\vspace{-0.1in}
\caption{Example of Code Evolution: From Raw Data to Theoretical Coding.}
\label{fig:Quali}
\end{figure}

During focused coding, related codes were grouped into higher-level categories, such as \textit{Empathy in AI Development}. The analysis progressed to theoretical coding, where connections between categories were established to construct a cohesive explanation of the observed phenomenon, grounded in participants’ experiences. 
Emerging themes were iteratively validated against the data, ensuring they accurately represented participants’ experiences. Figure \ref{fig:Quali} illustrates this process.

Throughout the analysis, memos played a key role in documenting emerging findings, tracking the evolution of codes, and exploring relationships between categories. These memos served as dynamic records of the analytical process, helping to clarify connections between codes and refine emerging themes. By continually revisiting and updating memos, we ensured consistency and transparency in data interpretation. Additionally, in accordance with the GT guidelines \citep{charmaz2014constructing, adolph2011using, stol2016grounded, ralph2020empirical}, memos were used not only to record evolving thoughts and interpretations but also to drive theoretical sensitivity by encouraging constant engagement with the data. They facilitated critical reflection on emerging concepts and their interrelations, ultimately shaping the direction of subsequent coding and theoretical development.

\subsection{Auditing}
To ensure rigor and transparency, we maintained a comprehensive audit trail throughout the study. This included detailed documentation of sampling decisions, coding practices, and memo development. Consistent with the guidelines \citep{charmaz2014constructing, adolph2011using, stol2016grounded, ralph2020empirical}, adjustments to the interview guide, informed by emerging data, were systematically recorded, such as refining questions to probe the implications of diversity. These iterative changes ensured that the data collection process remained responsive to the study's evolving findings. 

Later, member checking \citep{lincoln1988criteria, motulsky2021member, santos2017member} was employed to validate the accuracy and relevance of our findings, with selected participants reviewing and providing feedback on the emerging themes. This process not only confirmed alignment with their lived experiences but also offered an opportunity to refine our interpretations. For the member-checking process, we randomly selected participants from different projects and roles, ensuring a representative cross-section of perspectives. During these sessions, we presented the emerging themes and our understanding of the insights derived from the interviews \citep{santos2017member}. Participants were invited to comment, clarify, or challenge these interpretations, fostering a collaborative dialogue that strengthened the validity of our findings. Engaging with participants allowed us to refine and contextualize themes, ensuring they were firmly grounded in their narratives and reflected the complexities of team diversity in AI development. Additionally, peer debriefing sessions among the research team fostered discussions to evaluate interpretations and enhance the overall credibility of the findings.

\subsection{Ethics}
This study adhered to the ethical guidelines of the first author’s institution. Participants were fully informed of the research objectives, the voluntary nature of their involvement, and their right to withdraw at any time. Participants did not receive any financial or material compensation for their participation in the study.
 Informed consent was obtained before each interview, and all data were anonymized to protect participant confidentiality. Reflexivity and positionality \citep{de2024integrating} were also considered, with the research team discussing potential individual biases and their influence on data interpretation to safeguard the integrity of the findings.

\section{Findings} 
\label{sec:results}
This section presents the categories that emerged from the analysis, emphasizing how team diversity contributes to identifying and addressing bias in AI development. To situate the findings, we begin with demographic information about the participants, followed by a discussion of the categories. Table~\ref{tab:DiversityImpact} provides an overview of the key insights.

\subsection{Demographics}

As described in Section~\ref{collection}, we employed convenience sampling, snowball sampling, and theoretical sampling to recruit 25 professionals actively engaged in developing AI-powered systems. These participants were involved in diverse technologies, including deep learning neural networks, prediction models, LLMs, and computer vision applications for facial recognition. The sample demonstrated considerable diversity across gender, professional roles, educational backgrounds, professional experience, ethnicity, disability status, LGBTQIA+ representation, and neurodivergence, as shown in Table \ref{tab:Demographics} and summarized below:

\begin{itemize}
    \item \textbf{Gender and Representation:} The sample included a mix of participants who self-identified as men, women, and non-binary participants, reflecting varied gender identities and enriching discussions on inclusivity in team dynamics. While these represent three commonly recognized gender categories, we acknowledge that this does not encompass the full spectrum of gender identities. In particular, individuals from other gender identities often belong to hidden or extremely underrepresented populations in software engineering \citep{de2024hidden}, and without explicit self-identification, it is not possible to collect data from them. Specifically, our sample included 16 men, 8 women, and 1 non-binary participant. \\ 
    
    \item \textbf{Professional Roles and Expertise:} Participants represented a wide range of roles, from data scientists and designers to testers, programmers, researchers, and managers, providing varied experience and perspectives about the AI development lifecycle. In total, there were 10 data scientists, 5 designers, 4 testers, 3 programmers, 2 researchers, and 1 manager. \\
    
    \item \textbf{Educational and Professional Backgrounds:} The sample was composed by varied educational attainment, from bachelor’s degrees to PhDs, and professional experience ranging from early-career professionals to those with over a decade working in software development. Nine participants held bachelor’s degrees, 4 had postbaccalaureate certificates, 7 had master’s degrees, and 5 had PhDs. Regarding professional experience, 3 participants had 1–3 years, 11 had 3–5 years, 6 had 5–10 years, and 5 had more than 10 years. \\
    
    \item \textbf{Equity-Deserving Groups:} The inclusion of LGBTQIA+ individuals, neurodivergent participants, and professionals with disabilities ensured perspectives from equity-deserving groups, contributing to a holistic understanding of diversity’s role in AI development. The sample included 6 LGBTQIA+ participants, 4 neurodivergent participants, and 2 individuals with disabilities. \\
    
    \item \textbf{Ethnic and Cultural Perspectives:} While predominantly White, the sample also included participants who identified as mixed-race and Black, allowing for the consideration of cultural and racial perspectives within AI team dynamics. Specifically, 20 participants self-identified as White, 4 as mixed-race, and 1 as Black.
\end{itemize}

Participants were distributed across four ongoing AI projects in the company. While the individual served as the source of information in the interviews, our unit of analysis is the project. Participants were asked to reflect on their experiences as members of specific AI development teams, and their responses centered on how team diversity shaped project-related processes and outcomes. The focus was not on how diversity impacted them or their individual work, but rather on how diverse perspectives and interactions influenced tasks such as identifying bias or making design and implementation decisions. In this way, each participant’s narrative provided evidence about the functioning of the project team as a whole, grounded in real-world AI development work.

Diversity within our sample was emphasized intentionally, aligning with the study’s focus on equity and fairness, and supported by prior research on the importance of diverse team compositions in producing inclusive software outcomes~\citep{rodriguez2021perceived,albusays2021diversity}. The diversity across demographic, professional, and experiential dimensions was essential for capturing a broad range of perspectives. The inclusion of equity-deserving groups, such as LGBTQIA+ professionals, neurodivergent individuals, and those with disabilities, ensured that the findings reflected the lived experiences of underrepresented populations who are often disproportionately affected by the very systems they help build. These perspectives provided valuable information about how diversity influences fairness-oriented decision-making, the identification of biases, and the recognition of harmful outcomes. Furthermore, variation in roles, educational backgrounds, and levels of experience allowed for an examination of fairness concerns across different stages of the AI lifecycle, from requirements through testing.

\begin{table}
\centering
\caption{Demographics}
\renewcommand{\arraystretch}{1}
\label{tab:Demographics}
\begin{tabular}{llr}
\toprule
\multirow{2}{*}{ \textbf{Gender} } 
& Men & 16 individuals\\
& Women & 8 individuals\\ 
& Non-binary & 1 individual \\ \midrule 

\multirow{6}{*}{ \textbf{Role} } 
& Data Scientists & 10 individuals\\
& Designers & 5 individuals\\
& Programmers & 3 individuals\\
& Testers & 4 individuals\\
& Researchers & 2 individuals\\
& Managers & 1 individual \\ \midrule 

\multirow{4}{*}{ \textbf{Education} } 
& Bachelor's Degree & 9 individuals\\ 
& Postbaccalaureate & 4 individuals\\ 
& Master's Degree & 7 individuals\\ 
& PhD Degree & 5 individuals\\ \midrule 

\multirow{4}{*}{ \textbf{Experience} }  
& 1-3 years & 3 individuals\\ 
& 3-5 years & 11 individuals\\ 
& 5-10 years & 6 individuals\\ 
& More than 10 years & 5 individuals\\ \midrule 

\multirow{3}{*}{ \textbf{Ethnicity} } 
& White & 20 individuals\\
& Mixed-race & 4 individuals\\
& Black & 1 individual \\ \midrule 

\multirow{2}{*}{ \textbf{Disability} } 
& Without & 23 individuals\\
& With & 2 individuals \\ \midrule 

\multirow{2}{*}{ \textbf{LGBTQIA+} } 
& No & 19 individuals\\
& Yes & 6 individuals \\ \midrule 

\multirow{2}{*}{ \textbf{Neurodivergent} } 
& No & 21 individuals\\
& Yes & 4 individuals\\
\bottomrule
\end{tabular}
\end{table}

\subsection{Emergence of Themes}
Our findings yielded six categories that explain how team diversity supports the identification and mitigation of bias in AI systems. These categories were developed through an iterative coding process informed by Grounded Theory (see Section~\ref{analysis}). Line-by-line coding of transcripts generated initial codes such as \textit{multiple perspectives}, \textit{lived experience}, and \textit{context sensitivity}. Focused coding then grouped these into broader categories. For instance, codes related to context sensitivity, fostering empathy, and high sensitivity informed the category \textit{Bringing Empathy to AI Development}, while references to bias identification and cognitive bias reduction contributed to \textit{Using Diversity as a Safeguard Against Bias}. Similarly, codes such as multiple perspectives, different views, individual experience, and lived experience informed \textit{Diversifying Perspectives for Bias Identification}, while broadened thinking and diverse problem solving shaped \textit{Leveraging Diverse Expertise to Tackle Complex Bias}. Codes emphasizing contextual knowledge contributed to \textit{Promoting Inclusive and Fair Decision-Making}, and those highlighting enhanced understanding informed \textit{Addressing Systemic Discrimination}. Through constant comparison, we observed that practices such as incorporating domain expertise and lived experience were particularly relevant to addressing bias across different stages of development. The final categories include: \textit{Diversifying Perspectives for Bias Identification}, \textit{Bringing Empathy to AI Development}, \textit{Addressing Systemic Discrimination}, \textit{Promoting Inclusive and Fair Decision-Making}, \textit{Leveraging Diverse Expertise to Tackle Complex Bias}, and \textit{Using Diversity as a Safeguard Against Bias}. Collectively, these categories illustrate how diverse teams contribute to the recognition, mitigation, and prevention of bias throughout the development process.

\subsection{The Role of Team Diversity in Developing Unbiased AI Systems}
Team diversity contributes to the development of AI systems by supporting practices aimed at fairness and inclusiveness. Our findings indicate that diverse teams influence the development process through varied perspectives, empathetic considerations, attention to discrimination, improved decision-making, and the ability to address complex societal and multidisciplinary challenges. These roles are discussed in detail below, with supporting evidence from participant experiences summarized in Table~\ref{tab:DiversityImpact}.

\begin{itemize}
   \item \textbf{\textit{Diversifying Perspectives for Bias Identification:}} Team diversity supports the identification of different forms of bias, including racial, gender, and social biases, across various stages of AI system development. Varied perspectives allow potential issues to be recognized earlier, contributing to the design of systems that reflect a broader range of users and contexts. This aspect of diversity extends beyond addressing social inequality to include questioning oversimplified assumptions, narrow problem framing, and limited understanding of how systems are used in practice. Drawing on a wider set of lived experiences, disciplinary expertise, and contextual knowledge, diverse teams are able to highlight limitations in how data is collected, interpreted, and applied. Such teams are also more likely to ask whose behaviors are being encoded as “default,” which scenarios are prioritized, and whether critical use cases have been considered. In this way, diversity relates not only to ethical concerns but also to technical robustness, as it enables the identification of edge cases and contextual mismatches within the process. Several participants described this role of diversity in practice. For example, P11 noted how the presence of a deaf team member prompted a reassessment of interaction design, shifting accessibility from a peripheral concern to a reconsideration of underlying design assumptions. Similarly, P01 and P14 explained how gender-diverse teams surfaced embedded assumptions in algorithmic logic, such as defaults based on male experiences or overlooking alternative user flows. P09 and P17 highlighted that diverse perspectives facilitated more comprehensive evaluations of requirements and prototypes, making it possible to identify blind spots and anticipate usability challenges that might otherwise have been overlooked.
   
   \item \textbf{\textit{Bringing Empathy to AI Development:}} Diversity within teams was associated with fostering empathy, understood as an increased capacity to recognize and consider the experiences and challenges of underrepresented or marginalized groups. In the data, empathy was described as an important factor for anticipating how AI systems might be received and used by different social groups. Participants noted that in teams with less diversity, attention tended to be directed primarily toward technical aspects of system features, while social and emotional contexts of technology use could be overlooked. In contrast, diverse teams were described as more likely to reflect on how systems might affect people, particularly in situations involving sensitive attributes. Empathy was not framed as an innate trait but as a capacity that could be cultivated when individuals worked alongside teammates with diverse lived experiences. This process was linked to efforts to identify potential harms caused by AI models and to consider alternative design choices. For example, P06 highlighted the difficulty of practicing empathy when developers lacked awareness of possible harms, particularly if they were disconnected from affected communities. P14 noted that developers often focused on deadlines and technical outputs unless the social consequences of design decisions became personally evident. P17 described how diversity enabled teams to move beyond efficiency-driven development to engage with users’ lived experiences, which he referred to as understanding “the perception of the pain.” Similarly, P07 and P09 explained that collaborating with teammates from different backgrounds encouraged reflection on user realities and prompted questioning of default assumptions in system behavior.

    \item \textbf{\textit{Addressing Systemic Discrimination:}} Diversity within teams was described as contributing to the prevention and mitigation of systemic discrimination by broadening the range of perspectives considered in AI development. Including members from different racial, ethnic, gender, and socio-economic backgrounds was associated with a greater likelihood of recognizing both explicit forms of inequality, such as racial or gender exclusion, and more subtle forms embedded in model training and evaluation. Participants noted that many AI systems fail to adequately serve marginalized populations not because of deliberate intent, but because those affected by structural inequality were absent from the development process. In such cases, discriminatory behaviors could remain unchallenged and be treated as edge cases or statistical anomalies. Diverse teams were described as bringing awareness to these issues earlier, making it more likely that systemic harms were identified and addressed prior to deployment. For example, P05 explained that computer vision systems often struggled with the accurate interpretation of dark-skinned individuals, and that a lack of racial and cultural diversity within teams could delay both recognition and remediation of this problem. Other participants emphasized similar dynamics. Several participants noted that the absence of groups such as Black, Indigenous, people with disabilities, or LGBTQIA+ individuals in the early stages of development could lead to additional costs later in the process (e.g., addressing defects, managing negative publicity) and limit opportunities to design for inclusion. P01, P02, P13, and P22 emphasized that teams with limited diversity were more likely to treat dominant experiences as the norm and to overlook failures that had disproportionate effects on underrepresented communities.

    \item \textbf{\textit{Promoting Inclusive and Fair Decision-Making:}} Diversity was described as shaping decision-making across the development process, with implications for inclusiveness and fairness in AI systems. Rather than a one-time checkpoint, participants characterized fairness decision-making as an ongoing discussion of priorities, values, and trade-offs. Teams composed of individuals with varied backgrounds, domains, and lived experiences were more likely to consider how model outcomes might affect different groups of users and to question whose needs were being addressed. Such teams also reported engaging more frequently in collaborative adjustments to design goals, data strategies, and evaluation criteria to better reflect user diversity. For example, P03 emphasized the importance of early-stage decisions, explaining how intentional planning to include people with reduced mobility or visible conditions in their team led to fairer training sets in their project. P05 and P12 described how fairness considerations were strengthened when data and design criteria were intentionally diversified. P20 explained that decisions about system behavior became more robust in their project when informed by feedback from underrepresented groups and through the involvement of domain specialists in testing and development.
    
   \item \textbf{\textit{Leveraging Diverse Expertise to Tackle Complex Bias:}} Diversity was described as contributing to problem-solving by bringing a broad range of expertise into the development process, enabling teams to address complex bias-related challenges. The analysis indicated that such biases often stem from overlapping factors that cannot be resolved through isolated adjustments. Diverse teams were seen as better positioned to navigate this complexity by combining experiential knowledge and technical expertise, which supported the design of more contextually robust solutions. This process frequently involved questioning default assumptions and resisting simplified approaches, such as indiscriminately expanding datasets or applying uniform model parameters. For example, P05 explained how their team revised a dataset to include individuals with vitiligo and other visible skin conditions that standard models had failed to represent, highlighting the effort required to move beyond default demographic categories. P14 observed that common industry practices often reinforced prevailing societal norms rather than interrogating them. Participants P09, P18, and P19 emphasized the importance of involving people with deep domain knowledge or lived experience to strengthen teams’ ability to meet fairness guidelines.
    
    \item \textbf{\textit{Using Diversity as a Safeguard Against Bias:}} Diversity within teams was described as serving a protective role against bias by embedding a range of perspectives throughout the AI development process. Rather than depending solely on isolated checks or designated roles, structural diversity was associated with distributing responsibility for identifying and addressing bias across the team. Participants characterized this as creating conditions for ongoing ethical reflection and collective responsibility, making it more likely that assumptions and biased data would be questioned before becoming part of system logic. For example, P05 noted that fairness reviews were more substantive when conducted within diverse groups, as members contributed different experiences to assess potential harms. P22 observed that in the absence of real diversity, even structured ethics reviews risked becoming superficial. Similarly, P03, P21, and P23 described diversity as functioning like a \textit{bias buffer}, reducing the likelihood that unchecked assumptions would be embedded into the system.
\end{itemize}

\begin{table*}
\centering
\tiny
\caption{Team Diversity Role in AI Development}
\renewcommand{\arraystretch}{1.3}
\label{tab:DiversityImpact}
\begin{tabularx}{\linewidth}{p{1.9cm} p{2.7cm} X}
\toprule
\textbf{Diversity Role} & \textbf{Description} & \textbf{Evidence} \\ \midrule

\textbf{Diversifying Perspectives for Bias Identification} &  Team diversity supports the identification and mitigation of different biases in different phases of AI systems development. & "If we have an entire team of white people... there might not be someone with the sensitivity to notice that (racial) bias is occurring." (P02). \newline "If we didn't have individuals who are deaf involved in the process, for us, there will be a lot of effort and money spent to try to understand it [the problem]." (P05) \newline "I see the composition of the team as super important for building fair solutions, because it’s through diverse teams that we bring different perspectives." (P12) \\ \midrule

\textbf{Bringing Empathy to AI Development} & Team diversity fosters empathy, enabling teams to understand and consider the effects of biases on underrepresented groups. &  "Yes, I don’t know if I would call it empathy, but I think it would be a soft skill that some people should learn." (P01) \newline "I think that those who do not experience a particular context will not necessarily have the empathy, the sensitivity, or the care to check, test, and analyze whether a particular aspect is not being considered or is at risk of bias." (P02) \newline "Then having people from these groups becomes important, right? Simply due to empathy. Look, if I’m developing a product for myself, of course I will think about my own experience, right?" (P06) \\ \midrule

\textbf{Addressing Systemic Discrimination} & Diverse teams help to prevent and mitigate systemic discrimination, ensuring that AI systems do not perpetuate societal biases. & "If I don’t have LGBT people, if I don’t have Black people, if I don’t have Indigenous people, if I don’t have Indian people, then you start looking at it and say, wait, I’m not being represented here." (P05) \newline "Incorporating knowledge from Indigenous communities, Black communities, queer communities, and other groups into these AI systems, as well as professionalizing these minority groups so that they are represented in tech teams." (P22) \\ \midrule

\textbf{Promoting Inclusive and Fair Decision-Making} & Team diversity enhances decision-making in the software development process, ensuring that AI systems are designed to be inclusive, fair, and representative of various demographics. & "Having a person from the deaf community in the project completely changed the way we worked." (P10) \newline "If there's no diversity, decisions might favor certain groups." (P11) \newline "Diversity of thoughts can open more possibilities." (P17);  \\ \midrule

\textbf{Leveraging Diverse Expertise to Tackle Complex Bias} & Team diversity enriches problem-solving by bringing multiple viewpoints, leading to more innovative and comprehensive solutions to biases. & "The more diverse the team (...) we will be able to anticipate problems because many minds, especially with different experiences, think better than one person or a closed group." (P12) \newline "Yes, for sure. It's not just about understanding the problem, but I believe that having this plurality in the team—people from different origins, niches, and ways of thinking—can open up more possibilities." (P17) \\ \midrule

\textbf{Using Diversity as a Safeguard Against Bias} & Diverse teams serves as a safeguard against biases by providing multiple checks throughout AI system development. & "The existence of a culture of diversity in the company and within the team would be a first step in objectively questioning the development of potentially discriminatory features." (P21) \newline "Any team that lacks diversity in the people working on it has a greater chance of building systems with cognitive bias." (P23). \\ \bottomrule
\end{tabularx}
\end{table*}

Overall, our findings indicate that team diversity functions as more than a demographic characteristic within AI systems development. Participants described it as shaping how problems were framed, how solutions were designed, and how potential real-world consequences were considered. Across different roles such as bias identification, decision-making, and addressing complex challenges, diversity was associated with reflection, collaboration, and attention to fairness. In this way, diversity appeared not only to contribute to technical quality but also to support the development of AI systems that are more inclusive and context-aware.

\subsection{Team Diversity Shapes AI Development Processes}

The six categories described above represent roles that team diversity can take in the development of AI systems, from the perspective of software practitioners. Rather than functioning independently, these roles were often portrayed by practitioners as interconnected and reinforcing one another across different phases of development. For example, empathy was linked to recognizing systemic discrimination, which in turn supported more inclusive decision-making. Similarly, identifying bias through diverse perspectives was connected to preventive practices, such as embedding diversity as a safeguard. From practitioners’ accounts, these relationships suggested that diversity operates as part of an ongoing process of identifying, interpreting, and addressing bias, rather than as a single intervention or a static feature of team composition.

Our analysis also indicated that these roles frequently appeared in overlapping and iterative ways, according to practitioners. They described, for instance, how decision-making became more inclusive when teams had already developed habits of reflecting on user perspectives and anticipating varied needs. In such cases, inclusiveness was not reported as a separate stage but as part of how technical trade-offs and fairness were considered throughout development. Practitioners also connected the idea of diversity as a safeguard to early-stage practices such as identifying biased training data or questioning initial assumptions, suggesting that structural team diversity created conditions for continuous reflection rather than occasional oversight.

The categories also differed in how they were represented in practitioners’ accounts. Some, such as diversifying perspectives and fostering empathy, were described across multiple teams as enabling other practices. Others, such as using diversity as a safeguard, were more often framed in terms of long-term influence on team culture or organizational structure. This variation suggests that, from the perspective of practitioners, the roles of diversity are not uniform, and that particular aspects may be more or less prominent depending on team composition, the stage of development, and the AI application under consideration.

Finally, it is important to note that participants in our study did not emphasize negative aspects of team diversity, even though such challenges have been discussed in the literature, particularly in relation to conflicts and the complexity of decision-making processes such as reconciling differing viewpoints. The absence of such accounts in our data may be related to the framing of this study. Participants were invited to reflect on how team diversity influences bias identification and mitigation, two actions that are generally regarded as positive within responsible AI development. Therefore, practitioners may have been more inclined to highlight beneficial aspects of diversity. At this point, we understand that a study with a broader focus on team functioning, including interaction, communication, and coordination during AI development, might reveal experiences where team diversity presents challenges alongside benefits.

\section{Discussions}

We start these discussions by answering our research question, followed by a comparison between our findings and related work. We finish by presenting implications to research and practice, as well as discussing the limitations of our research.

\subsection{Answering the Research Question}

Based on the key themes identified in our analysis and supported by the evidence collected from the interviews, we address the general research question: \textbf{\textit{What is the role of team diversity in the identification and mitigation of biases in AI system development?}} Overall, our findings indicate that diversity was not regarded as a background feature of team composition but as an active factor shaping how biases and fairness were discussed in AI system development. Our analysis suggests that team diversity played a formative role in how bias was identified, interpreted, and addressed throughout the development process. This diversity influenced how problems were defined, which data was considered relevant, and the goals established for the development of a fair software product. This characteristic enabled teams to anticipate fairness concerns that might remain overlooked in more homogeneous groups.

Six interconnected roles captured how diversity contributed to bias identification and mitigation. Varied perspectives helped uncover design assumptions and blind spots in data use (\textit{Diversifying Perspectives for Bias Identification}). Working with colleagues from different backgrounds fostered empathy, prompting reflection on how systems might affect users differently, particularly those from marginalized groups (\textit{Bringing Empathy to AI Development}). This awareness supported the detection of exclusionary practices and the questioning of normalized defaults (\textit{Addressing Systemic Discrimination}). These reflections informed more deliberate discussions about fairness in requirements, data selection, and evaluation (\textit{Promoting Inclusive and Fair Decision-Making}). When bias was complex, involving overlapping technical and social factors, combining diverse expertise provided teams with a stronger basis for developing solutions (\textit{Leveraging Diverse Expertise to Tackle Complex Bias}). Finally, diversity was described as a structural condition that embedded fairness into daily routines, distributing responsibility across the team rather than limiting it to isolated checks or individual roles (\textit{Using Diversity as a Safeguard Against Bias}).

\subsection{Comparing Findings}\label{comparing}
Previous research has shown that AI systems can replicate societal inequalities present in their training data, leading to discriminatory outcomes, particularly for marginalized groups \citep{ferrara2023fairness, lee2020artificial, allen2020artificial, brun2018software, zhang2021ignorance, paez2021negligent, kelly2023algorithmic}. Our findings align with the ongoing discussion on the importance of identifying strategies to mitigate such inequalities. However, unlike most studies in the software engineering field \citep{nguyen2024literature, desoftware}, our results focus on social aspects, with particular attention to the role of team diversity. Exploring diversity in this way highlights its potential relevance for detecting different forms of bias linked to social factors. This perspective contributes to ongoing debates suggesting that addressing algorithmic discrimination requires attention not only to technical approaches but also to the human and organizational dimensions of AI development.

When considering the roots of bias, such as cognitive, design, historical, and model biases \citep{santos2024preliminary}, our analysis suggests that team diversity may help mitigate these challenges. Participants described ways in which diverse perspectives contributed to identifying and addressing different forms of bias across the development process. Specifically, team diversity was reported to play a role in recognizing cognitive and societal biases, while also supporting practices that addressed requirement, training, and testing biases through more inclusive decision-making. These findings complement existing literature that has predominantly emphasized technical aspects of datasets and models, suggesting that diverse teams may also provide a basis for questioning assumptions and refining development practices with attention to fairness.

On a different note, the literature on team diversity in software engineering has often investigated managerial dimensions, such as productivity and teamwork \citep{menezes2018diversity, wickramasinghe2015diversity, verwijs2023double}. Our study broadens this perspective by focusing on the contribution of team diversity to AI development processes. This aligns with emerging discussions in related fields where diversity is increasingly recognized as relevant for the success of AI projects. In this respect, our results address a gap in the software engineering literature and indicate that the intersection of diversity and AI development warrants further investigation.

At the same time, prior studies have also documented challenges associated with diversity. Communication barriers, limited feedback, and cultural misunderstandings have been reported in multicultural agile teams, particularly in early stages before shared practices are established \citep{welsch2024navigating}. Research on collaborative programming tasks has suggested that diversity can introduce discomfort, anxiety, and hesitance in team formation, even when task outcomes are positive \citep{mason2024diversity}. Studies of globally distributed teams have also observed that demographic diversity may lead to relational conflict and that certain diversity dimensions, such as gender, can increase tension without necessarily improving team performance \citep{verwijs2023double, wickramasinghe2015diversity}. In contrast, participants in our study did not report interpersonal conflict or diversity-related tension. This absence may reflect characteristics of the organizational settings, the maturity of the teams involved, or limitations of the method, as participants were not anonymous and may have provided socially desirable responses \citep{kwak2021does}. It is also possible that, since bias mitigation is typically regarded as a positive goal in AI development, participants maintained a positive view when reflecting on the theme. Nonetheless, participants were asked about the aspects of team diversity using a neutral framing, which suggests that the positive experiences may have been more salient in the contexts studied. Further research employing alternative designs could provide additional insights into dynamics that were not captured in our data.

Our findings also contribute to discussions on how team diversity may address the root causes of bias in AI systems \citep{desoftware}. Participants described how working in diverse teams helped them revisit assumptions and consider user needs more broadly (\textit{Diversifying Perspectives for Bias Identification}), which relates to challenges of \textit{cognitive bias} and \textit{requirements bias} \citep{zhang2021ignorance, desoftware}. Accounts of empathy for users who belong to groups not present in the team, fostered through exposure to different lived experiences, were linked to anticipating harms and reconsidering design choices (\textit{Bringing Empathy to AI Development}), as well as addressing concerns around \textit{design bias} and \textit{societal bias} \citep{brun2018software, hort2024bias, desoftware}. In some cases, participants noted that including professionals from equity-deserving groups enabled recognition of exclusionary patterns in data and task framing (\textit{Addressing Systemic Discrimination}), which relates to issues of \textit{historical bias} \citep{allen2020artificial, hoffmann2019fairness, desoftware}. Discussions about data strategy and evaluation practices (\textit{Promoting Inclusive and Fair Decision-Making} and \textit{Leveraging Diverse Expertise to Tackle Complex Bias}) showed how diverse expertise informed model training and validation, addressing challenges of \textit{training bias} and \textit{model bias} \citep{ferrara2023fairness, chen2024fairness, desoftware}. Finally, participants described fairness as a shared responsibility within diverse teams rather than assigned to specific individuals, which aligns with efforts to reduce \textit{testing bias} and narrow validation strategies (\textit{Using Diversity as a Safeguard Against Bias}) \citep{galhotra2017fairness, soremekun2022software, desoftware}.


\subsection{Implications}\label{implications}
Our findings have implications for research, industry practice, and broader societal contexts. By investigating how team diversity influences bias identification and mitigation in AI development, our study contributes to ongoing discussions on fairness and inclusivity in software engineering. These implications extend beyond technical considerations, pointing to the ways in which diverse teams may shape the design and deployment of AI systems with greater attention to equity across different settings.

\subsubsection{Implications for Research}
Our findings open an avenue for investigating issues such as bias in AI and algorithmic discrimination from a socially grounded perspective. This approach aligns with broader societal challenges and highlights the need for a more comprehensive understanding of how social dynamics, particularly software team characteristics, shape AI systems. By shifting attention beyond the purely technical aspects that have traditionally dominated software engineering research, this study highlights the importance of considering the socio-technical implications of AI development and the integration of social factors into discussions of algorithmic fairness. The qualitative design of this study, based on interviews and interactions with practitioners, offers an alternative methodological approach to exploring research questions within software engineering for AI. This methodology allows for in-depth exploration of real-world experiences and social dynamics, generating perspectives that may not be accessible through other approaches, e.g., experiments and benchmarks. Engaging directly with practitioners provided context-specific accounts of how team diversity influences the development of AI systems, emphasizing the intersection of social and technical factors in addressing algorithmic fairness.

This work also points to several directions for future research. One important area is the exploration of different dimensions of diversity, such as gender, ethnicity, and sexual orientation, and their distinct roles in AI system development. Although practitioners in this study referred to multiple forms of diversity, the evidence was not sufficient to examine each dimension in depth. Future research could investigate these factors individually to develop a more detailed understanding of their contributions to bias identification and mitigation in AI development. Other opportunities for research include examining how organizational structures and cultural contexts influence the effectiveness of diverse teams in reducing AI biases, as well as studying how diversity-focused strategies can be systematically incorporated into software engineering processes.

\subsubsection{Implications for Practice}
Our findings provide insights that may be relevant for industry practitioners working on AI system development. By highlighting the role of team diversity in identifying and mitigating biases, this research suggests ways in which practitioners might refine their approaches to bias detection and mitigation. Software teams can benefit from being more intentional about incorporating diverse perspectives, recognizing that diversity is not only an ethical consideration but also a practical strategy for developing AI systems that attend to fairness. In this regard, the following recommendations are drawn from our findings:

\begin{itemize}
\item \textbf{Foster diverse team composition across technical roles and responsibilities.} Participants reported that diverse backgrounds helped identify bias in training data, requirements, and model outputs. This finding is consistent with prior research indicating that varied perspectives support early problem identification \citep{welsch2024navigating, devathasan2025empathy}. AI teams may therefore benefit from including a mix of disciplinary, demographic, and experiential backgrounds across all stages of development.

\item \textbf{Create structured spaces for empathy and experience sharing.} Participants noted that empathy influenced more inclusive design decisions. Research has shown that empathy and emotional regulation are important for effective collaboration in diverse teams \citep{devathasan2025empathy}. Activities such as reflective design practices or inclusive design storytelling could help integrate empathy into everyday development routines.

\item \textbf{Support inclusive and participatory decision making.} Participants emphasized that inclusive discussions improved fairness-oriented decisions. In line with literature connecting psychological safety to collaborative performance \citep{verwijs2023double}, organizations may consider formalizing participatory processes such as fairness checkpoints or collaborative retrospectives, creating spaces where assumptions can be questioned and results validated.

\item \textbf{Incorporate multi-perspective testing practices.} Diversity was described as functioning as a safeguard during development. The literature also highlights the value of feedback from a range of stakeholders \citep{wickramasinghe2015diversity}. Structured fairness testing that draws on perspectives from underrepresented groups may strengthen evaluation practices.

\item \textbf{Ensure inclusion is sustained through team continuity and leadership support.} While participants in our study did not report conflict, other studies have described interpersonal tension in diverse teams \citep{mason2024diversity, welsch2024navigating}. Investments in long-term team stability and leadership training may help foster trust and inclusive cultures that support collaboration in AI development.
\end{itemize}

Integrating these strategies into everyday development practices could help software organizations strengthen fairness-related outcomes while also contributing to the overall quality and accountability of AI systems.

\subsubsection{Implications for Society} 
Our findings relate to the broader societal relevance of team diversity in AI development. By showing how diversity influences the identification and mitigation of bias, our study demonstrates the connection between technical practices in software engineering and persistent social challenges such as discrimination based on race, gender, or socioeconomic status. Our findings reinforce the importance of human-centered approaches to AI design, ensuring that technologies are not only innovative but also attentive to fairness and equity. Addressing these concerns is important for reducing the risk of reproducing systemic discrimination and for promoting outcomes that are more inclusive. Our findings also indicate that social factors shape both the effectiveness and the fairness of AI systems. Encouraging diversity in AI development teams may contribute to technologies that better reflect the varied needs of society, support underrepresented communities, and foster greater public trust in AI.

\subsection{Threats to Validity}
Our findings should be interpreted in light of the limitations of the Grounded Theory methodology used in this study \citep{charmaz2014constructing, ralph2020empirical}. This qualitative approach prioritizes criteria such as credibility, resonance, usefulness, and contribution over traditional quantitative measures such as validity and generalizability. To enhance credibility, we established a clear chain of evidence linking participants’ experiences to our conclusions. We sought to include diverse perspectives by involving participants of different genders, ethnicities, and sexual orientations. Resonance was supported through iterative refinement of findings with participants. The usefulness of the study lies in its focus on team diversity in relation to AI bias, while its contribution stems from bringing social dynamics into discussions of AI systems development.

As with most qualitative studies \citep{franklin2001reliability, verdecchia2023threats}, our work is not intended to provide broad statistical generalization. Instead, we conducted detailed interviews with professionals until data saturation was reached, producing insights that are transferable to comparable contexts rather than universally applicable. We also provided descriptive information about participants to support re-analysis and transferability. While we identified possible relationships between team diversity and improved bias detection in AI, these are based on qualitative data and should be interpreted cautiously when applied to other settings.

Although the study was conducted within a single company, several factors increase the relevance of the findings. The unit of analysis was the software project rather than the company itself. Projects were developed for different clients, with varied technical domains and development goals. Clients included both multinational corporations and regional organizations, and development teams worked in close collaboration with client professionals, reflecting global software development practices. This cross-organizational collaboration adds to the contextual richness and supports the transferability of the findings.

We also acknowledge the possibility of social desirability bias \citep{kwak2021does}. Because participants shared their experiences in semi-structured interviews without anonymity, responses may have been influenced by the presence of the researcher. Although participants were asked neutrally about several aspects of team diversity, challenges commonly reported in other studies, such as communication issues, interpersonal tension, or conflict, did not emerge in our data. This contrasts with studies that relied on anonymous surveys or indirect methods, where participants may have felt more comfortable reporting sensitive experiences. It is therefore possible that our findings reflect more socially acceptable narratives, particularly given the fairness-focused framing of the study.

Finally, we recognize potential threats to theoretical validity, defined here as the extent to which the relationships we construct among concepts represent participants’ intended meanings \citep{petersen2013worldviews}. Because our study explored how team diversity relates to bias mitigation in AI systems, interpretive reasoning was required to construct links between concepts such as empathy, fairness, and representation. To strengthen theoretical validity, we used strategies such as constant comparison, peer debriefing, and member checking to ensure that conceptual integration was traceable to participants’ accounts. Nonetheless, our theorizing is contextually situated and not exhaustive. Rather than proposing a generalizable model, our aim was to provide a plausible and useful explanation grounded in the perspectives of those we interviewed.

\section{Conclusion}
In this study, we investigated the role of team diversity in detecting and mitigating biases in AI systems, with a focus on how diverse team compositions may influence the development of fairer software. By considering a range of backgrounds, perspectives, and lived experiences within development teams, we identified ways in which these elements were reported to contribute to recognizing and addressing biases, including those related to race, gender, and social factors. Our analysis of real-world AI development projects indicated patterns suggesting that team diversity contributed to surfacing biases that might otherwise have remained unnoticed.

Our study contributes to the body of knowledge in software engineering by identifying six categories through which team diversity was described as influencing AI development: \textit{(a) Diversifying Perspectives for Bias Identification}, \textit{(b) Bringing Empathy to AI Development}, \textit{(c) Addressing Systemic Discrimination}, \textit{(d) Promoting Inclusive and Fair Decision-Making}, \textit{(e) Leveraging Diverse Expertise to Tackle Complex Bias}, and \textit{(f) Using Diversity as a Safeguard Against Bias}. These categories illustrate how diversity was viewed as supporting bias identification and mitigation across different stages of the AI development lifecycle. Unlike research that has primarily emphasized managerial aspects of diversity, our findings bring attention to the social and practical roles that diverse teams may play in AI development.

The implications of these findings are relevant for research, practice, and society. For the research community, this study suggests the value of integrating social factors alongside technical considerations in order to develop a more comprehensive understanding of AI bias and fairness. For practitioners, the results point to the potential benefits of intentionally cultivating diverse development teams, recognizing diversity not only as an ethical consideration but also as a practical means of identifying and addressing biases. These insights may inform diversity initiatives connected to specific stages of AI development, such as bias detection, data preparation, and system testing. Finally, at a societal level, as AI systems are increasingly deployed in critical domains such as healthcare, education, and public services, ensuring fairness and inclusivity becomes essential. Diverse teams were described as bringing perspectives that help design AI technologies more aligned with the needs of different communities, which may reduce the risk of reinforcing existing biases and contribute to greater trust in AI.

For future work, we plan to build on these findings by investigating how specific dimensions of diversity, including gender, ethnicity, sexual orientation, and neurodivergence, influence bias identification and mitigation in AI systems. This involves studying how these dimensions shape collaboration, decision-making, and the broader development process. We also intend to incorporate insights from gray literature and practitioner-oriented sources, such as developer forums and industry reports, to complement our empirical findings and enhance their applicability. In addition, we aim to employ a range of empirical methods to design strategies for integrating diversity-focused practices into software engineering workflows. The goal of this effort is to generate approaches that are both generalizable and practical, providing guidance for practitioners seeking to strengthen team effectiveness while supporting fairness in AI development.

\section {Data Availability}
In this \url{https://figshare.com/s/f0344a831d50167dc649}, a set of anonymized quotations extracted from the interviews is available. We have selected quotations that exclude direct references to participants' names, co-workers, projects, or companies. As a result, some participants from our sample may not have associated quotations in the spreadsheet. Additionally, some quotations might read awkwardly as they were directly translated from the participants' native languages.

\section* {Conflict of Interest}

The authors declare that they have no conflict of interest.

\section* {Funding}
This work was supported by the Natural Sciences and Engineering Research Council of Canada (NSERC), Discovery Grant RGPIN-2024-06260, and by Alberta Innovates through the Advance Program, Project Number 24506125.

\section* {Ethical approval}
This study adhered to the ethical guidelines of the first author’s institution under REB24-0357. 

\section* {Informed Consent}
Participants were fully informed of the research objectives, the voluntary nature of their involvement, and their right to withdraw at any time. Informed consent was obtained before each interview, and all data were anonymized to protect participant confidentiality.

\section* {Author Contributions}
The authors declare that they have contributed equally to this paper.


%

\bibliographystyle{spbasic}      
\bibliography{bib}   

@book{charmaz2014constructing,
  title={Constructing grounded theory},
  author={Charmaz, Kathy},
  year={2014},
  publisher={sage}
}

@inproceedings{stol2016grounded,
  title={Grounded theory in software engineering research: a critical review and guidelines},
  author={Stol, Klaas-Jan and Ralph, Paul and Fitzgerald, Brian},
  booktitle={Proceedings of the 38th International conference on software engineering},
  pages={120--131},
  year={2016}
}

@book{glaser1978theoretical,
  title={Theoretical sensitivity},
  author={Glaser, Barney G},
  year={1978},
  publisher={University of California,}
}

@article{rodriguez2021perceived,
  title={Perceived diversity in software engineering: a systematic literature review},
  author={Rodr{\'\i}guez-P{\'e}rez, Gema and Nadri, Reza and Nagappan, Meiyappan},
  journal={Empirical Software Engineering},
  volume={26},
  number={5},
  pages={1--38},
  year={2021},
  publisher={Springer}
}

@article{albusays2021diversity,
  title={The diversity crisis in software development},
  author={Albusays, Khaled and Bjorn, Pernille and Dabbish, Laura and Ford, Denae and Murphy-Hill, Emerson and Serebrenik, Alexander and Storey, Margaret-Anne},
  journal={IEEE Software},
  volume={38},
  number={2},
  pages={19--25},
  year={2021},
  publisher={IEEE}
}

@article{baltes2022sampling,
  title={Sampling in software engineering research: A critical review and guidelines},
  author={Baltes, Sebastian and Ralph, Paul},
  journal={Empirical Software Engineering},
  volume={27},
  number={4},
  pages={1--31},
  year={2022},
  publisher={Springer}
}

@article{lee2018detecting,
  title={Detecting racial bias in algorithms and machine learning},
  author={Lee, Nicol Turner},
  journal={Journal of Information, Communication and Ethics in Society},
  volume={16},
  number={3},
  pages={252--260},
  year={2018},
  publisher={Emerald Publishing Limited}
}

@article{soremekun2022software,
  title={Software fairness: An analysis and survey},
  author={Soremekun, Ezekiel and Papadakis, Mike and Cordy, Maxime and Traon, Yves Le},
  journal={arXiv preprint arXiv:2205.08809},
  year={2022}
}

@article{chen2024fairness,
  title={Fairness testing: A comprehensive survey and analysis of trends},
  author={Chen, Zhenpeng and Zhang, Jie M and Hort, Max and Harman, Mark and Sarro, Federica},
  journal={ACM Transactions on Software Engineering and Methodology},
  volume={33},
  number={5},
  pages={1--59},
  year={2024},
  publisher={ACM New York, NY}
}

@article{starke2022fairness,
  title={Fairness perceptions of algorithmic decision-making: A systematic review of the empirical literature},
  author={Starke, Christopher and Baleis, Janine and Keller, Birte and Marcinkowski, Frank},
  journal={Big Data \& Society},
  volume={9},
  number={2},
  pages={20539517221115189},
  year={2022},
  publisher={SAGE Publications Sage UK: London, England}
}

@inproceedings{brun2018software,
  title={Software fairness},
  author={Brun, Yuriy and Meliou, Alexandra},
  booktitle={Proceedings of the 2018 26th ACM joint meeting on european software engineering conference and symposium on the foundations of software engineering},
  pages={754--759},
  year={2018}
}

@inproceedings{zhang2021ignorance,
  title={" Ignorance and Prejudice" in Software Fairness},
  author={Zhang, Jie M and Harman, Mark},
  booktitle={2021 IEEE/ACM 43rd International Conference on Software Engineering (ICSE)},
  pages={1436--1447},
  year={2021},
  organization={IEEE}
}

@inproceedings{verma2018fairness,
  title={Fairness definitions explained},
  author={Verma, Sahil and Rubin, Julia},
  booktitle={Proceedings of the international workshop on software fairness},
  pages={1--7},
  year={2018}
}

@inproceedings{galhotra2017fairness,
  title={Fairness testing: testing software for discrimination},
  author={Galhotra, Sainyam and Brun, Yuriy and Meliou, Alexandra},
  booktitle={Proceedings of the 2017 11th Joint meeting on foundations of software engineering},
  pages={498--510},
  year={2017}
}

@online{mari2020libras,
  title={Lenovo And Brazilian Innovation Hub CESAR Create Sign Language “Translator” For Hearing People With AI},
  author={Mari, A},
  year={2023},
  publisher={Forbes},
  url={https://www.forbes.com/sites/angelicamarideoliveira/2023/08/09/lenovo-and-brazilian-innovation-hub-cesar-create-sign-language-translator-for-hearing-people-with-ai/},
  lastaccessed = "Jul. 5, 2024"
}

@article{boch2022ethical,
  title={Ethical artificial intelligence in paediatrics},
  author={Boch, Samantha and Sezgin, Emre and Linwood, Simon Lin},
  journal={The Lancet Child \& Adolescent Health},
  volume={6},
  number={12},
  pages={833--835},
  year={2022},
  publisher={Elsevier}
}

@article{hoffmann2019fairness,
  title={Where fairness fails: data, algorithms, and the limits of antidiscrimination discourse},
  author={Hoffmann, Anna Lauren},
  journal={Information, Communication \& Society},
  volume={22},
  number={7},
  pages={900--915},
  year={2019},
  publisher={Taylor \& Francis}
}

@article{kirat2023fairness,
  title={Fairness and Explainability in Automatic Decision-Making Systems. A challenge for computer science and law},
  author={Kirat, Th and Tambou, Olivia and Do, Virginie and Tsouki{\`a}s, Alexis},
  journal={EURO journal on decision processes},
  volume={11},
  pages={100036},
  year={2023},
  publisher={Elsevier}
}

@article{hort2024bias,
  title={Bias mitigation for machine learning classifiers: A comprehensive survey},
  author={Hort, Max and Chen, Zhenpeng and Zhang, Jie M and Harman, Mark and Sarro, Federica},
  journal={ACM Journal on Responsible Computing},
  volume={1},
  number={2},
  pages={1--52},
  year={2024},
  publisher={ACM New York, NY}
}

@inproceedings{allen2020artificial,
  title={Artificial Intelligence: the right to protection from discrimination caused by algorithms, machine learning and automated decision-making},
  author={Allen, Robin and Masters, Dee},
  booktitle={ERA Forum},
  volume={20},
  number={4},
  pages={585--598},
  year={2020},
  organization={Springer}
}

@article{rodriguez2023lgbtq,
  title={LGBTQ incorporated: YouTube and the management of diversity},
  author={Rodriguez, Julian A},
  journal={Journal of Homosexuality},
  volume={70},
  number={9},
  pages={1807--1828},
  year={2023},
  publisher={Taylor \& Francis}
}

@article{ryan2021digital,
  title={How digital beauty filters perpetuate colorism: an ancient form of prejudice about skin color is flourishing in the modern internet age},
  author={Ryan-Mosly, T},
  journal={Technology Review},
  volume={15},
  year={2021}
}

@inproceedings{wu2020gender,
  title={Gender classification and bias mitigation in facial images},
  author={Wu, Wenying and Protopapas, Pavlos and Yang, Zheng and Michalatos, Panagiotis},
  booktitle={Proceedings of the 12th ACM Conference on Web Science},
  pages={106--114},
  year={2020}
}

@article{fountain2022moon,
  title={The moon, the ghetto and artificial intelligence: Reducing systemic racism in computational algorithms},
  author={Fountain, Jane E},
  journal={Government Information Quarterly},
  volume={39},
  number={2},
  pages={101645},
  year={2022},
  publisher={Elsevier}
}

@article{owens2020those,
  title={Those designing healthcare algorithms must become actively anti-racist},
  author={Owens, Kellie and Walker, Alexis},
  journal={Nature medicine},
  volume={26},
  number={9},
  pages={1327--1328},
  year={2020},
  publisher={Nature Publishing Group US New York}
}

@inproceedings{santos2023perspective,
  title={The Perspective of Software Professionals on Algorithmic Racism},
  author={Santos, Ronnie de Souza and de Lima, Luiz Fernando and Magalhaes, Cleyton},
  booktitle={Proceedings of the 17th ACM/IEEE International Symposium on Empirical Software Engineering and Measurement},
  year={2023}
}

@article{paez2021negligent,
  title={Negligent algorithmic discrimination},
  author={P{\'a}ez, Andr{\'e}s},
  journal={Law \& Contemp. Probs.},
  volume={84},
  pages={19},
  year={2021},
  publisher={HeinOnline}
}

@article{garcia2024algorithmic,
  title={Algorithmic discrimination in the credit domain: what do we know about it?},
  author={Garcia, Ana Cristina Bicharra and Garcia, Marcio Gomes Pinto and Rigobon, Roberto},
  journal={AI \& SOCIETY},
  volume={39},
  number={4},
  pages={2059--2098},
  year={2024},
  publisher={Springer}
}

@article{kelly2023algorithmic,
  title={Algorithmic discrimination at work},
  author={Kelly-Lyth, Aislinn},
  journal={European Labour Law Journal},
  volume={14},
  number={2},
  pages={152--171},
  year={2023},
  publisher={SAGE Publications Sage UK: London, England}
}

@article{serna2022sensitive,
  title={Sensitive loss: Improving accuracy and fairness of face representations with discrimination-aware deep learning},
  author={Serna, Ignacio and Morales, Aythami and Fierrez, Julian and Obradovich, Nick},
  journal={Artificial Intelligence},
  volume={305},
  pages={103682},
  year={2022},
  publisher={Elsevier}
}

@article{nguyen2024literature,
  title={From Literature to Practice: Exploring Fairness Testing Tools for the Software Industry Adoption},
  author={Nguyen, Thanh and de Lima, Luiz Fernando and Badassarre, Maria Teresa and Santos, Ronnie de Souza},
  journal={18th ACM/IEEE International Symposium on Empirical Software Engineering and Measurement},
  year={2024}
}

@article{limante2024bias,
  title={Bias in Facial Recognition Technologies Used by Law Enforcement: Understanding the Causes and Searching for a Way Out},
  author={Limantė, Agnė},
  journal={Nordic Journal of Human Rights},
  volume={42},
  number={2},
  pages={115--134},
  year={2024},
  publisher={Taylor \& Francis}
}

@incollection{laupman2022biased,
  title={Biased algorithms and the discrimination upon immigration policy},
  author={Laupman, Clarisse and Schippers, Laurianne-Marie and Papal{\'e}o Gagliardi, Marilia},
  booktitle={Law and artificial intelligence: regulating AI and applying AI in legal practice},
  pages={187--204},
  year={2022},
  publisher={Springer}
}

@article{adams2020diversity,
  title={The diversity crisis of software engineering for artificial intelligence},
  author={Adams, Bram and Khomh, Foutse},
  journal={IEEE Software},
  volume={37},
  number={5},
  pages={104--108},
  year={2020},
  publisher={IEEE}
}

@article{ferrara2023fairness,
  title={Fairness and bias in artificial intelligence: A brief survey of sources, impacts, and mitigation strategies},
  author={Ferrara, Emilio},
  journal={Sci},
  volume={6},
  number={1},
  pages={3},
  year={2023},
  publisher={MDPI}
}

@book{lee2020artificial,
  title={Artificial intelligence in daily life},
  author={Lee, Raymond ST},
  year={2020},
  publisher={Springer}
}

@article{bigman2023algorithmic,
  title={Algorithmic discrimination causes less moral outrage than human discrimination.},
  author={Bigman, Yochanan E and Wilson, Desman and Arnestad, Mads N and Waytz, Adam and Gray, Kurt},
  journal={Journal of Experimental Psychology: General},
  volume={152},
  number={1},
  pages={4},
  year={2023},
  publisher={American Psychological Association}
}

@inproceedings{dehal2024exposing,
  title={Exposing Algorithmic Discrimination and Its Consequences in Modern Society: Insights from a Scoping Study},
  author={Dehal, Ramandeep Singh and Sharma, Mehak and de Souza Santos, Ronnie},
  booktitle={Proceedings of the 46th International Conference on Software Engineering: Software Engineering in Society},
  pages={69--73},
  year={2024}
}

@article{santos2024preliminary,
  title={Preliminary Insights on Industry Practices for Addressing Fairness Debt},
  author={de Souza Santos, Ronnie and de Lima, Luiz Fernando and Baldassarre, Maria Teresa and Spinola, Rodrigo},
  journal={18th ACM/IEEE International Symposium on Empirical Software Engineering and Measurement},
  year={2024}
}

@article{schwarting2022organization,
  title={Why Organization Matters in “Algorithmic Discrimination”},
  author={Schwarting, Rena and Ulbricht, Lena},
  journal={KZfSS K{\"o}lner Zeitschrift f{\"u}r Soziologie und Sozialpsychologie},
  volume={74},
  number={Suppl 1},
  pages={307--330},
  year={2022},
  publisher={Springer}
}

@inproceedings{pieterse2006software,
  title={Software engineering team diversity and performance},
  author={Pieterse, Vreda and Kourie, Derrick G and Sonnekus, Inge P},
  booktitle={Proceedings of the 2006 annual research conference of the South African institute of computer scientists and information technologists on IT research in developing countries},
  pages={180--186},
  year={2006}
}

@inproceedings{menezes2018diversity,
  title={Diversity in software engineering},
  author={Menezes, {\'A}lvaro and Prikladnicki, Rafael},
  booktitle={Proceedings of the 11th International workshop on cooperative and human aspects of software engineering},
  pages={45--48},
  year={2018}
}

@article{wickramasinghe2015diversity,
  title={Diversity in team composition, relationship conflict and team leader support on globally distributed virtual software development team performance},
  author={Wickramasinghe, Vathsala and Nandula, Sahan},
  journal={Strategic Outsourcing: An International Journal},
  volume={8},
  number={2/3},
  pages={138--155},
  year={2015},
  publisher={Emerald Group Publishing Limited}
}

@article{verwijs2023double,
  title={The Double-Edged Sword of Diversity: How Diversity, Conflict, and Psychological Safety Impact Software Teams},
  author={Verwijs, Christiaan and Russo, Daniel},
  journal={IEEE Transactions on Software Engineering},
  year={2023},
  publisher={IEEE}
}

@inproceedings{gila2014impact,
  title={Impact of personality and gender diversity on software development teams' performance},
  author={Gila, Abdul Rehman and Jaafa, Jafreezal and Omar, Mazni and Tunio, Muhammad Zahid},
  booktitle={2014 International Conference on Computer, Communications, and Control Technology (I4CT)},
  pages={261--265},
  year={2014},
  organization={IEEE}
}

@article{desoftware,
  title={Software Fairness Debt: Building a Research Agenda for Addressing Bias in AI Systems},
  author={de Souza Santos, Ronnie and Fronchetti, Felipe and Freire, S{\'a}vio and Spinola, Rodrigo},
  journal={ACM Transactions on Software Engineering and Methodology},
  year={2024},
  publisher={ACM New York, NY}
}

@article{adolph2011using,
  title={Using grounded theory to study the experience of software development},
  author={Adolph, Steve and Hall, Wendy and Kruchten, Philippe},
  journal={Empirical Software Engineering},
  volume={16},
  pages={487--513},
  year={2011},
  publisher={Springer}
}

@techreport{lincoln1988criteria,
  title={Criteria for Assessing Naturalistic Inquiries as Reports},
  author={Lincoln, Yvonna S and Guba, Egon G},
  institution={ERIC},
  year={1988}
}

@article{motulsky2021member,
  title={Is member checking the gold standard of quality in qualitative research?},
  author={Motulsky, Sue L},
  journal={Qualitative Psychology},
  volume={8},
  number={3},
  pages={389},
  year={2021},
  publisher={Educational Publishing Foundation}
}

@inproceedings{santos2017member,
  title={Member checking in software engineering research: Lessons learned from an industrial case study},
  author={Santos, Ronnie ES and Magalhaes, Cleyton VC and Da Silva, Fabio QB},
  booktitle={2017 ACM/IEEE International Symposium on Empirical Software Engineering and Measurement (ESEM)},
  pages={187--192},
  year={2017},
  organization={IEEE}
}

@article{de2024integrating,
  title={Integrating Positionality Statements in Empirical Software Engineering Research},
  author={de Sousa, Breno Felix and Santos, Ronnie de Souza and Gama, Kiev},
  journal={arXiv preprint arXiv:2412.06567},
  year={2024}
}

@article{franklin2001reliability,
  title={Reliability and validity in qualitative research},
  author={Franklin, Cynthia and Ballan, Michelle},
  journal={The handbook of social work research methods},
  volume={4},
  number={273-292},
  year={2001}
}

@article{verdecchia2023threats,
  title={Threats to validity in software engineering research: A critical reflection},
  author={Verdecchia, Roberto and Engstr{\"o}m, Emelie and Lago, Patricia and Runeson, Per and Song, Qunying},
  journal={Information and Software Technology},
  volume={164},
  pages={107329},
  year={2023},
  publisher={Elsevier}
}

@online{botocario,
  title={Beauty Meets AI: World’s First Prototype Smart Lipstick®},
  author={Grupo Boticario},
  year={2025},
  publisher={Grupo Boticario},
  url={https://www.grupoboticario.com.br/en/midia/smart-lipstick-ai-prototype/},
  lastaccessed = "Sep. 27, 2025"
}

@inproceedings{de2024hidden,
  title={Hidden populations in software engineering: Challenges, lessons learned, and opportunities},
  author={de Souza Santos, Ronnie and Gama, Kiev},
  booktitle={Proceedings of the 1st IEEE/ACM International Workshop on Methodological Issues with Empirical Studies in Software Engineering},
  pages={58--63},
  year={2024}
}

@inproceedings{welsch2024navigating,
  title={Navigating cultural diversity: barriers and benefits in multicultural agile software development teams},
  author={Welsch, Daniel and Burk, Luisa and M{\"o}tefindt, David and Neumann, Michael},
  booktitle={Proceedings of the 39th ACM/SIGAPP Symposium on Applied Computing},
  pages={818--825},
  year={2024}
}

@article{mason2024diversity,
  title={Diversity’s Double-Edged Sword: Analyzing Race’s Effect on Remote Pair Programming Interactions},
  author={Mason, Shandler A and Kuttal, Sandeep Kaur},
  journal={ACM transactions on software engineering and methodology},
  volume={34},
  number={1},
  pages={1--45},
  year={2024},
  publisher={ACM New York, NY}
}

@article{kwak2021does,
  title={When does social desirability become a problem? Detection and reduction of social desirability bias in information systems research},
  author={Kwak, Dong-Heon Austin and Ma, Xiao and Kim, Sumin},
  journal={Information \& Management},
  volume={58},
  number={7},
  pages={103500},
  year={2021},
  publisher={Elsevier}
}

@article{devathasan2025empathy,
  title={Empathy, self-determination and motivation: moderating diversity for enhanced performance in software development teams},
  author={Devathasan, Kezia and Arony, Nowshin Nawar and Murphy-Hill, Emerson and Damian, Daniela},
  journal={Empirical Software Engineering},
  volume={30},
  number={3},
  pages={82},
  year={2025},
  publisher={Springer}
}

@inproceedings{petersen2013worldviews,
  title={Worldviews, research methods, and their relationship to validity in empirical software engineering research},
  author={Petersen, Kai and Gencel, Cigdem},
  booktitle={2013 joint conference of the 23rd international workshop on software measurement and the 8th international conference on software process and product measurement},
  pages={81--89},
  year={2013},
  organization={IEEE}
}

@article{pfeiffer2023algorithmic,
  title={Algorithmic fairness in AI: an interdisciplinary view},
  author={Pfeiffer, Jella and Gutschow, Julia and Haas, Christian and M{\"o}slein, Florian and Maspfuhl, Oliver and Borgers, Frederik and Alpsancar, Suzana},
  journal={Business \& Information Systems Engineering},
  volume={65},
  number={2},
  pages={209--222},
  year={2023},
  publisher={Springer}
}

@inproceedings{giannopoulos2024fairness,
  title={Fairness in AI: challenges in bridging the gap between algorithms and law},
  author={Giannopoulos, Giorgos and Psalla, Maria and Kavouras, Loukas and Sacharidis, Dimitris and Marecek, Jakub and Matilla, Germ{\'a}n M and Emiris, Ioannis},
  booktitle={2024 IEEE 40th International Conference on Data Engineering Workshops (ICDEW)},
  pages={217--225},
  year={2024},
  organization={IEEE}
}


\appendix
\begin{table}
\centering
\caption{Interview Script: Exploring Team Diversity in AI Development}
\tiny
\renewcommand{\arraystretch}{1.3}
\label{tab:InterviewScript}
\begin{tabular}{p{2cm}p{9cm}}
\toprule
\textbf{Topic} & \textbf{Questions and Follow-ups} \\ \midrule

\textbf{Introduction} & 
\textbf{Main question:} Tell me about your day-to-day work with developing AI systems. \newline
\textbf{Probe:}
\begin{itemize}
    \item What type of software do you develop?
    \item What is your main role in the team?
\end{itemize} \\ \midrule

\textbf{Team Composition} & 
\textbf{Main question:} Tell me about the composition of your team. \newline
\textbf{Probe:}
\begin{itemize}
    \item How diverse would you say your team is?
    \item Is your team diverse in terms of technical characteristics? 
    \item Is your team diverse in terms of individual personal characteristics?    
\end{itemize} \\ \midrule

\textbf{Team Diversity and AI Development} & 
\textbf{Main question:} Considering the type of software you develop, comment about any impact of having a diverse team working with you. \\ \midrule

\textbf{Bias and Fairness} & 
\textbf{Main question:} Considering that biases can distort AI outcomes, potentially leading to errors or harmful results for specific groups of users, what do you do to identify and mitigate such biases? \newline
\textbf{Probe:}
\begin{itemize}
    \item How does your team handle these biases?
    \item How does the diversity of your team impact bias identification?
    \item Please share an example of when team diversity helped identify a bias.
    \item What procedures does your team use to address identified biases?
    \item Please share an example of when team diversity helped mitigate a bias.
\end{itemize} \\ \midrule

\textbf{Specific Bias Types} & 
\textbf{Main question:} Based on your experience, what types of biases can team diversity help identify or mitigate in AI system development? \newline
\textbf{Probe:} I will list various forms of bias documented in the literature. Please discuss whether team diversity contributed to addressing these biases in your project: \newline
\begin{itemize}
    \item \textbf{Cognitive Bias}: Developers' judgments or preconceived attitudes influence decisions, potentially perpetuating inequalities.
    \item \textbf{Design Bias}: Preconceived notions or technical preferences influence architectural and design decisions, reinforcing inequalities.
    \item \textbf{Historical Bias}: Historical events or contexts influence decisions and outcomes, perpetuating past discrimination.
    \item \textbf{Model Bias}: Systematic errors or inaccuracies in computational models result in distorted predictions, reinforcing inequalities.
    \item \textbf{Requirements Bias}: Bias in project specifications or user requirements influences feature definitions, perpetuating discrimination.
    \item \textbf{Social Bias}: Discriminatory practices embedded in societal norms or structures influence software outcomes, perpetuating inequalities.
    \item \textbf{Testing Bias}: Lack of diversity-focused testing strategies results in overlooked behaviors that perpetuate inequalities.
    \item \textbf{Training Bias}: Inaccuracies in training data lead to discriminatory results, reinforcing inequalities in software applications.
\end{itemize} \\ \midrule

\textbf{Soft Skills for AI Development} & 
\textbf{Main question:} What soft skills are necessary for professionals developing AI systems, considering the role of team diversity? \newline
\textbf{Probe:}
\begin{itemize}
    \item How do these skills impact the team?
    \item How do these skills impact the work?
    \item How do these skills impact the development of AI systems?
\end{itemize} \\ 
\end{tabular}
\end{table}

\end{document}